\documentclass[useAMS,usenatbib]{mn2e} 
\usepackage{graphicx}

\title[Orbital Period Glitches in XTE~J1710-281]{Eclipse Timings of the LMXB XTE~J1710-281: Orbital Period Glitches}

\author[Chetana Jain and Biswajit Paul]{Chetana Jain$^{1}$\thanks{E-mail:
chetanajain11@gmail.com (CJ); bpaul@rri.res.in (BP)} and Biswajit Paul$^{2}$\\
$^{1}$Hans Raj College, University of Delhi, Delhi 110007, India\\
$^{2}$Raman Research Institute, Sadashivnagar, C. V. Raman Avenue, Bangalore 560080, India}
\begin{document}

\date{ }

\pagerange{\pageref{firstpage}--\pageref{lastpage}} \pubyear{ }

\maketitle

\label{firstpage}

\begin{abstract}

We present an X-ray eclipse timing analysis of the transient low mass X-ray binary XTE~J1710-281. We report observations of 57 complete X-ray eclipses, spread over more than a decade of observations, made with the proportional counter array detectors aboard the $RXTE$ satellite. Using the eclipse timing technique, we have derived a constant orbital period of 0.1367109674 (3) d, during the period from MJD 52132 upto MJD 54410; and 1$\sigma$ limits of -1.6 $\times$ 10 $^{-12}$ d d$^{-1}$ and 0.2 $\times$ 10 $^{-12}$ d d$^{-1}$, on the period derivative, $\dot{P}_{orb}$. This puts constraints on the minimum timescale of secular orbital period evolution (${P}_{orb}/\dot{P_{orb}}$) of 2.34 $\times$ 10$^{8}$ yr for a period decay and 18.7 $\times$ 10$^{8}$ yr for a period increase. We also report detection of two instances of discontinuity in the mid-eclipse time, one before and one after the above MJD range. These results are interpreted as three distinct epochs of orbital period in XTE~J1710-281. We have put lower limits of 1.4 ms and 0.9 ms on orbital period change ($\Delta P_{orb}$) at successive epochs. The detection significance of the two orbital period glitches are 11$\sigma$ and 4$\sigma$ respectively. The sudden changes in orbital period is very similar in nature to that observed in EXO 0748-676, though their magnitude is much smaller in XTE J1710--281.

\end{abstract}


\section{Introduction}

XTE J1710--281 is a transient Low Mass X-ray Binary (LMXB) which was discovered in 1998 by the Rossi X-ray Timing Explorer ($RXTE$), and is likely to be associated with the ROSAT source 1RXS J171012.3-280754 \citep{Markwardt98}. It is a highly variable source and several bursts have been reported to occur \citep{Markwardt01,Galloway08}. The system has an orbital period of 3.28 hr \citep{Markwardt01} and the light curve shows dipping phenomena which could be due to occultations in the outer regions of the accretion disk as seen in many other high inclination LMXBs \citep{White82}. 

XTE J1710--281 is a poorly studied LMXB. Being in a binary system, the orbital period of XTE J1710--281 is expected to change, due to redistribution of the angular momentum arising from interaction between the components of the binary system. The orbit can evolve due to various mechanisms, such as, mass transfer within the system due to Roche lobe overflow, tidal interaction between the components of the binary system, gravitational wave radiation, magnetic braking, \citep{Rappaport83, Hurley02} and X-ray irradiated wind outflow \citep{Ruderman89}. The measurement of orbital period derivative and hence the orbital evolution is therefore, important to understand the physical processes occuring in the  system. 

Orbital evolution can be determined by several techniques. Measurement of pulse arrival time delay, is one of the methods used to determine the evolution of the binary orbits \citep{Deeter91, Paul04, Paul07}. The method of pulse folding and $\chi^{2}$ maximization with a varying orbital ephemeris, has also been used to determine the orbital evolution \citep{Paul02, Jain07}. Both these methods are well established and statistically at par. However, to determine the pulse arrival times, the data-length for each sample should be kept small so as to avoid significant smearing of pulse phase due to orbital motion. In the case of faint sources, or in the case of observation made with a small photon collection area (e.g. with focussing optics), a small integration time may be insufficient for measurement of pulse arrival time. In such a case, maximisation of pulse detection by varying the orbital ephemeris is an effective technique as the entire data set is used together \citep{Paul02}. Eclipse timing is another technique, used to determine the orbital evolution of LMXBs \citep{Wolff09, Jain10}. In some sources, in the absence of pulsations or eclipses, orbital period derivative has been measured with some stable orbital intensity modulation features (4U 1820-30: \citet{Chou01}; Cyg X-3: \citet{Singh02}).

XTE J1710--281 is one of the very few LMXBs, where full, sharp X-ray eclipses have been observed. The other systems being, EXO 0748--676 \citep{Wolff02}, GRS J1747--312 \citep{Zand03}, MXB 1658--298 \citep{Cominsky89} and AX J1745.6--2901 \citep{Maeda96}. Among these sources, EXO 0748--676 \citep{Parmar86, Wolff02} is the only system in which a large number of eclipses have been timed with high accuracy \citep{Wolff09}. In case of GRS J1747--312 and AX J1745.6--2901, the eclipse duration is too long ($\sim$43 minutes and $\sim$23 minutes, respectively, \citet{Zand03, Porquet07}) to carry out monitoring measurements. AX J1745.6--2901 is located near a bright source, which puts strong constraints on the timing analysis. The fifth source, MXB 1658--298 has been mostly in inactive state, since its discovery \citep{Wachter00, Oosterbroek01}. The LMXB XTE J1710--281 has a fairly small eclipse duration and has been persistently active since its discovery. This makes it an ideal source to investigate the orbital evolution.

In this work, we have used the sharp eclipses of XTE J1710--281 as timing markers to determine the orbital period of the system. By measuring the mid-eclipse times, over a long time baseline of $\sim$11 years, we can determine the change in the orbital period and hence estimate the orbital evolution of the system. 

\section{Observations and Analysis}

Data for the present analysis were obtained from observations made with the Proportional Counter Array (PCA) on board the Rossi X-ray Timing Explorer (\emph{RXTE}) satellite \citep{Bradt93}. The \emph{RXTE}-PCA consists of an array of five collimated xenon/methane multianode proportional counter units (PCU) with a total photon collection area of 6500 cm$^{2}$ \citep{Jahoda96, Jahoda06}, depending on the number of PCUs ON. The entire analysis was done using \textsc{ftools} from the astronomy software package \textsc{heasoft}-ver 6.10. The PCA data collected in the event mode and the Good Xenon mode, were used to generate the light curves, using the \textsc{ftool}-\textsc{seextrct}. The analysis was done in the energy band 2$-$20 keV. The background was estimated using the \textsc{ftool}-\textsc{pcabackest}. Faint source model was taken from the \emph{RXTE} website (http://heasarc.gsfc.nasa.gov/docs/xte/pca$_{-}$news.html). Thereafter, the background subtracted light curves were barycenter corrected using the \textsc{ftool}-\textsc{faxbary}.

We have analyzed all the \emph{RXTE}-PCA archival data, which covered a full X-ray eclipse. However, since this is a bursting source, we ignored few eclipses, where bursts occurred close to the ingress and egress phase of the eclipse. From data spread over $\sim$11 years (1999-2010), we have found 57 complete eclipses. Table 1 gives the observation IDs of all the 57 eclipses, observed with \emph{RXTE}-PCA. In Figure 1, we have shown a sample background subtracted light curve of XTE J1710-281 (Obs ID: 91045-01-01-13), binned with 2 seconds and including an eclipse lasting for $\sim$ 420 s, excluding the ingress and egress phase. 

\begin{figure}
% \centering
\includegraphics[height=3.5in, width=2.7in, angle=-90]{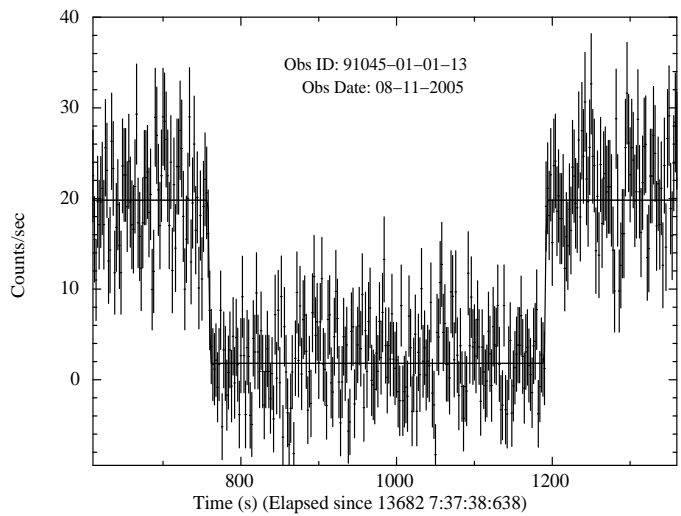}
\caption{A sample of 2$-$20 keV, background subtracted light curve of XTE J1710-281, binned with 2 s and including an eclipse lasting for about 428 s. The line represents the best fit five-parameter model for the light curve. The best fit had a $\chi^{2}$ of 381 for 375 d.o.f. }
\end{figure}

For most of the observations, apart from a few type-I X-ray bursts, the out-of-eclipse count rate did not seem to have any significant variability. Therefore, the variable components of the light curve around the eclipse phase are, the pre-ingress (C$_{pre-ingress}$), eclipse (C$_{eclipse}$), and post-egress (C$_{post-egress}$) count rates; the ingress ($\Delta t_{in} $) and egress ($ \Delta t_{eg} $) duration, the eclipse duration ($ \Delta t_{ecl} $) and the mid-eclipse time. Considering all the components to be freely variable, we first fitted a seven-parameter ramp and step model to the eclipse phase (similar to \citet{Wolff09}). It was found that the value of C$_{pre-ingress}$ and C$_{post-egress}$ were similar and the eclipse ingress and egress duration were also similar withing errors. The parameter space, thus got reduced to five: the pre-ingress and the post-egress count rate (C$_{pre-in-post-eg}$), the eclipse count rate, the ingress and egress duration ($ \Delta t_{in_eg} $), the eclipse duration ($ \Delta t_{ecl} $), and the mid-eclipse time. The average eclipse duration in XTE~J1710-281 is $\sim$ 420 s; and for the model fitting, $\sim$ 150 seconds of data was taken before and after the eclipse phase. It was also seen that the error in the mid eclispe time measurement was smaller when the five-parameter model was used. The best fit model for the eclipse profile in Figure 1, is shown with a solid line. The best fit had a $ \chi^{2} $ of 381 for 375 degrees of freedom.  

All the 57 X-ray eclipse light curves were fitted with a five-parameter ramp function, as described above. The mid-eclipse times and the corresponding 1$\sigma$ errors were determined. The results are given in Table 1. The errors in the mid-eclipse times vary between 0.000006 d to 0.000064 d, i.e. 0.5 s to 5.5 s. This is mainly due to difference in the relative count rates of the source and the background, and partly due to the number of detectors ON. The orbit numbers are with respect to the first eclipse detected in the \emph{RXTE}-PCA data. We fitted a constant and a linear model to the eclipse measurements between MJD 52132 - 54410. The results of the fit are given in Table 2. We obtained an orbital period of 0.1367109674 (3) d (epoch MJD 51250.924540 (4)) and have derived 1$\sigma$ limits of 0.2 $\times$ 10 $^{-12}$ d d$^{-1}$ and -1.6 $\times$ 10 $^{-12}$ d d$^{-1}$, on the orbital period derivative ($\dot{P}_{orb}$). The $\chi^{2}$ of the fit was 74 for 51 d.o.f. Before and after the above mentioned MJD range, we found shifts in the mid-eclipse times. We refer to these shifts as three epochs in the orbital period.  

\begin{figure}
\centering
\includegraphics[height=3.5in, width=2.7in, angle=-90]{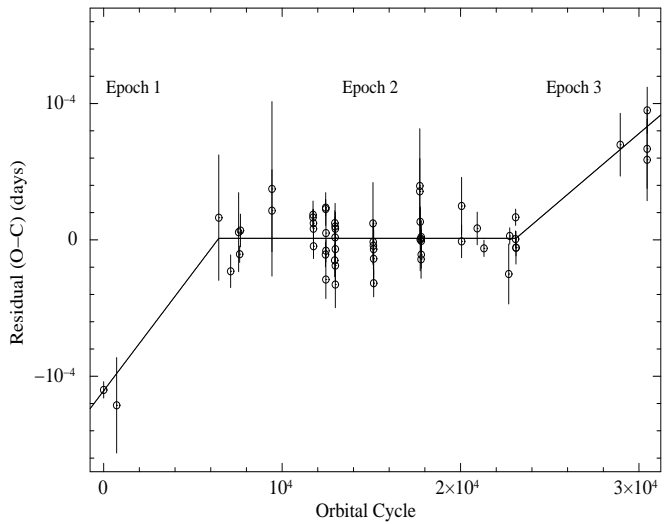}
\caption{The observed minus calculated times (residuals, O-C) for eclipses observed in XTE J1710-281 during 1999-2010, obtained from $RXTE$-PCA observations. The O-C variation is plotted as a function of orbital cycle. The three distinct orbital period epochs are mentioned.}
\end{figure}

Figure 2 shows the $``$observed minus calculated" (O - C) diagram for all the eclipse measurements of XTE J1710-281, obtained after subtracting the linear component obtained from epoch 2. The three different epochs of orbital period is evident. It is obvious from the figure that a polynomial function consisting of linear (P$_{orb}$), quadratic ($\dot{P_{orb}}$), cubic ($\ddot{P_{orb}}$) etc terms cannot be fitted to the observed dataset. A piecewise linear function could be more appropriate. But, there are few observations in epoch 1 and 3, hence one cannot determine the orbital period during epoch-1 and epoch-3 with very high accuracy. However, from our observations, we have put lower limits on orbital period changes of $\Delta P$ = 1.4 ms (1.7 $\times$ 10 $^{-8}$ d) between epoch-1 and epoch-2; and a $\Delta P$ = 0.9 ms (1.1 $\times$ 10 $^{-8}$ d) between epoch-2 and epoch-3. The detection significance of the two orbital period glitches are 11$\sigma$ and 4$\sigma$ respectively.

We have also created average eclipse profiles for the three epochs by combining all the corresponding eclipse light curves. The eclipse light curves were co-added using the orbital period and epoch given in Table 2. The folded profiles are shown in the top panel of Figure 3. The best fit ramp function is also shown for the eclipse profile of epoch 2. The best fit had a $ \chi^{2} $ of 689 for 650 degrees of freedom (d.o.f). The bottom panels of the same figure shows the eclipse ingress and egress profiles. It is clear from Figure 3 that the time of ingress and egress of eclipse, during the three epochs is shifted in phase while the eclipse duration has remained nearly same.

\begin{figure*}
\centering
\includegraphics[height=5.2in, width=2.7in, angle=-90]{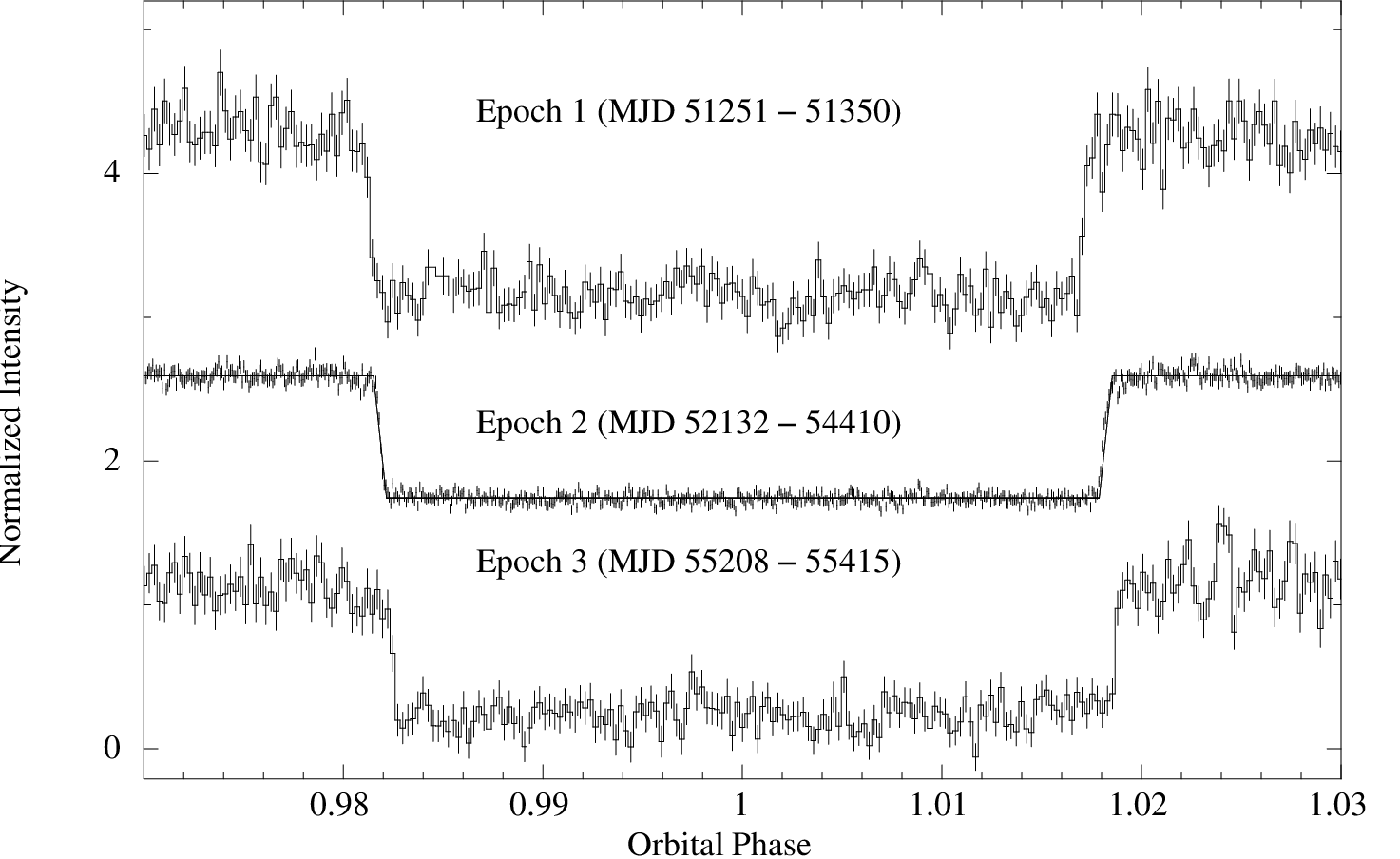}
\includegraphics[height=3.2in, width=2.7in, angle=-90]{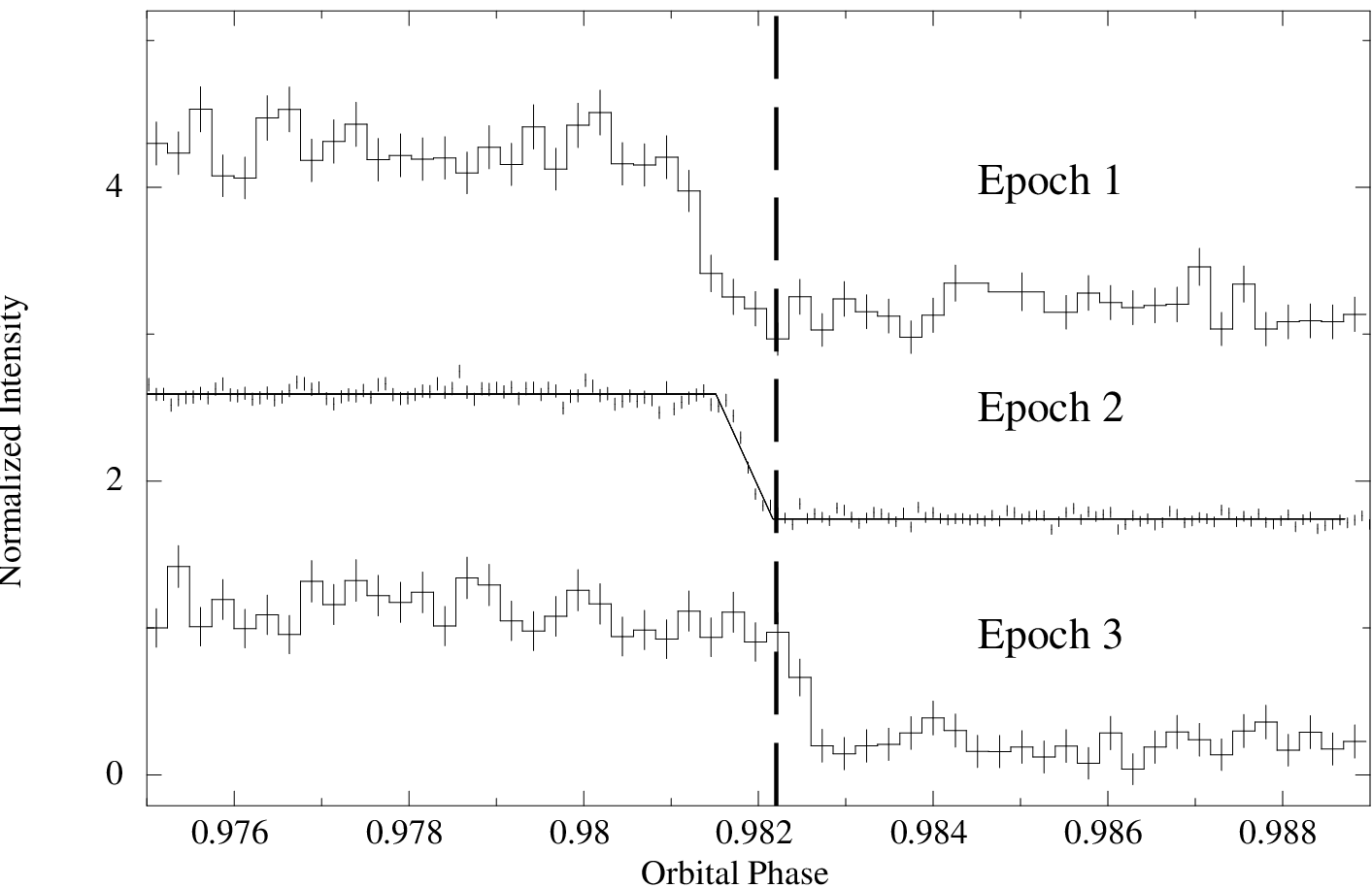}
\includegraphics[height=3.2in, width=2.7in, angle=-90]{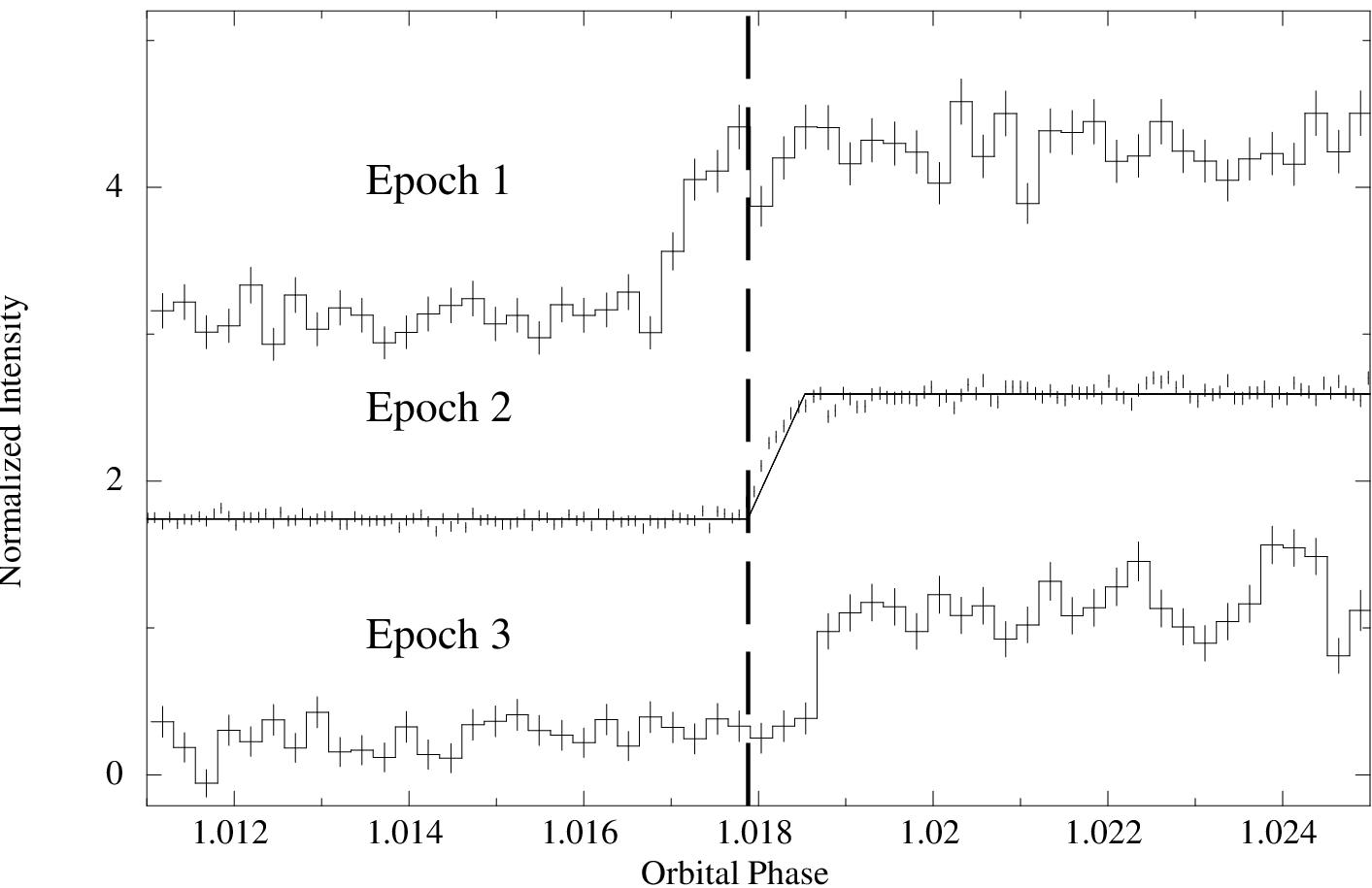}
\caption{A sample of 2$-$20 keV folded light curves of XTE J1710-281. The light curves were folded with a period of 0.1367109674 d at an epoch of MJD 51250.924540. The normalised intensities during epoch-1 and epoch-2 have been rescaled, by adding constant numbers to the curves. The solid line in the middle light curve (epoch 2) shows the best fit five-parameter model to the X-ray eclipse. The best fit had a reduced $\chi^{2}$ of 689 for 650 d.o.f. The top panel shows the complete eclipse, whereas, the bottom panels show the ingress and egress of the eclipse phase.}
\end{figure*}

\begin{table}
\centering
\begin{minipage}{140mm}
\caption{Mid Eclipse Time measurements of XTE J1710-281.}
\begin{tabular}{@{}lccc@{}}
\hline
\hline
$RXTE$		&	Orbital	&	Mid eclipse time 	&	1$\sigma$ Uncertainty \\
Observation ID	&	Cycle	&	MJD (d)			&	(d)\\
\hline
40407-01-03-00	&   	  1	&	 51251.061141	&       0.000006\\
40135-01-40-00	&    	728 	&        51350.450003  	&       0.000035\\
60049-01-01-03  &  	6452 	&        52132.983718  	&       0.000046\\
60049-01-03-00	&  	7118 	& 	 52224.033183 	&       0.000012\\
60049-01-04-00  &  	7567 	&        52285.416436 	&       0.000029\\
60049-01-05-00  &  	7623	&        52293.072234 	&       0.000006\\
60049-01-06-00  &  	7667	&        52299.087534 	&       0.000012\\
60049-01-07-010 &    	9431	&        52540.245711 	&       0.000064\\
60049-01-07-01  &  	9432	&        52540.382406 	&       0.000030\\
80045-01-01-00	&   	11732	&        52854.817626 	&       0.000006\\
80045-01-01-01	&  	11733	&        52854.954339	&       0.000010\\
80045-01-01-03	&  	11756	&        52858.098681 	&       0.000009\\
80045-01-01-04 	&  	11757	&        52858.235396 	&       0.000010\\
80045-01-01-05 	&  	11760	&        52858.645512 	&       0.000009\\
80045-01-02-00 	&  	12432	&        52950.515276 	&       0.000009\\
80045-01-02-01 	&  	12444	&        52952.155841 	&       0.000012\\
80045-01-02-03 	&  	12445	&        52952.292553 	&       0.000006\\
80045-01-02-02 	&  	12454	&        52953.522899 	&       0.000014\\
80045-01-02-04 	&  	12455	&        52953.659644 	&       0.000019\\
80045-01-02-05 	&  	12456	&        52953.796342 	&       0.000008\\
80045-01-03-00 	&  	12947 	&        53020.921420  	&       0.000012\\
80045-01-03-01 	&  	12957	&        53022.288557 	&       0.000009\\
80045-01-03-02 	& 	12970	&        53024.065797 	&       0.000017\\
80045-01-03-03 	&  	12971 	&        53024.202500 	&       0.000015\\
80045-01-03-04 	&  	12982	&        53025.706300 	&       0.000012\\
80045-01-03-05 	&  	12983	&        53025.843038	&       0.000012\\
80045-01-03-06 	&  	12984	&        53025.979734 	&       0.000008\\
80045-01-03-07 	&  	12985	&        53026.116419 	&       0.000017\\
80045-01-04-00 	&  	15096	&        53314.713316 	&       0.000030\\
80045-01-04-01 	&  	15107	&        53316.217123 	&       0.000013\\
80045-01-04-02 	&  	15115	&        53317.310808 	&       0.000017\\
80045-01-04-05 	&  	15129	&        53319.224759 	&       0.000021\\
80045-01-04-06 	&  	15130	&        53319.361463 	&       0.000021\\
80045-01-04-07 	&  	15131	&        53319.498156 	&       0.000010\\
91045-01-01-00 	&  	17718	&        53673.169500 	&       0.000042\\
91045-01-01-01 	&  	17719	&        53673.306207	&       0.000024\\
91045-01-01-02 	&  	17740	&        53676.177102	&       0.000023\\
91045-01-01-03 	&  	17741	&        53676.313826	&       0.000011\\
91045-01-01-13 	&  	17785	&        53682.329096	&       0.000009\\
91045-01-01-14 	&  	17786	&        53682.465792 	&       0.000014\\
91045-01-01-15 	&  	17797	&        53683.969629 	&       0.000012\\
91045-01-01-16 	&  	17798	&        53684.106327 	&       0.000010\\
91045-01-01-17 	&  	17799	&        53684.243048 	&       0.000008\\
91018-01-01-00 	&  	20059	&        53993.209860 	&       0.000021\\
91018-01-01-00 	&  	20060	&        53993.346545 	&       0.000012\\
91018-01-02-00	&  	20934	&        54112.831940 	&       0.000012\\
91018-01-03-01 	&  	21314	&        54164.782093 	&       0.000006\\
93052-01-01-01 	&  	22700	&        54354.263475 	&       0.000022\\
91018-01-07-04 	&  	22758	&        54362.192739 	&       0.000006\\
91018-01-08-01 	&  	23086	&        54407.033934 	&       0.000006\\
91018-01-08-01	&  	23087	&        54407.170661 	&       0.000006\\
91018-01-08-03 	&  	23107	&        54409.904858 	&       0.000012\\
91018-01-08-03 	& 	23108	&        54410.041569 	&       0.000006\\
94314-01-01-00 	&  	28952	&        55208.980538 	&       0.000023\\
94314-01-06-00	&  	30454	&        55414.320400  	&       0.000030\\
94314-01-06-00 	&  	30455	&        55414.457119  	&       0.000029\\
94314-01-06-01 	&  	30461 	&        55415.277413  	&       0.000017\\
\hline
\end{tabular}
 \end{minipage}
\end{table}

\begin{table}
\centering
\begin{minipage}{140mm}
\caption{Orbital ephemerides of XTE~J1710-281.}
\begin{tabular}{l r }
\hline
\hline
Parameter					&	Best fit value from present analysis\\
\hline
T$_{0}$ (MJD)					&	51250.924540 (4)\\
P$_{orb}$ (d)					&	0.1367109674 (3)\\
$\dot{P}_{orb}$	(10 $^{-12}$ d d$^{-1}$)	&	-1.6 $\leq$ $\dot{P}_{orb}$  $\leq$ 0.2 \\
${P}_{orb}/\dot{P_{orb}}$ (10 $^{8}$ yr) 	&	-2.34 $\leq$ ${P}_{orb}/\dot{P_{orb}}$ $\leq$ 18.7\\
\hline
\end{tabular}
 \end{minipage}
\end{table}

\section{Discussion}

Measurements of change in orbital period of an LMXB, is a crucial diagnostic to understand the accretion processes occuring in a binary system; and their effect on the system parameters. And the orbital period can be very well determined by time connecting a stable fiducial marker in the light curve of the binary system. We have analyzed 57 full X-ray eclipses of XTE J1710-281, observed by the $RXTE$ satellite. The observations cover more than 30000 binary orbits spread over $\sim$11 years. A five-component model was fit to each eclipse profile and the mid-eclipse times and the corresponding errors determined. We have determined an orbital period of 0.1367109674 (3) d and limits on the period derivative of -1.6 $\times$ 10 $^{-12}$ d d$^{-1}$ and 0.2 $\times$ 10 $^{-12}$ d d$^{-1}$, i.e., limits on the timescale of secular orbital period evolution, ${P}_{orb}/\dot{P_{orb}}$, of 2.34 $\times$ 10 $^{8}$ yr for a period decay and 18.7 $\times$ 10 $^{8}$ yr for a period increase. respectively, during the period from MJD 52132 to MJD 54410.

The variation in the orbital ephemerides of XTE~J1710-281, is significantly different from that seen in most of the other LMXB systems, such as, 4U 1820-303 \citep{Chou01}, SAX~J1808.4-3658 \citep{Jain07}, Her X-1 \citep{Paul07}, X~2127+119 \citep{Homer98} and 4U 1822-37 \citep{Jain10}. During the period from MJD 52132 to MJD 54410, the limits on the orbital period derivative of XTE~J1710-281, is more than an order of magnitude smaller than those measured in the other LMXBs (4U 1820-303 (3.5 $\times$ 10 $^{-8}$ yr$^{-1}$), SAX~J1808.4-3658 (1.3 $\times$ 10 $^{-8}$ yr$^{-1}$), Her X-1 (-2 $\times$ 10 $^{-7}$ yr$^{-1}$), X~2127+119 (9 $\times$ 10 $^{-7}$ yr$^{-1}$) and 4U 1822-37 (2.0 $\times$ 10 $^{-7}$ yr$^{-1}$)). Outside the above MJD range, the observed trend in the residual (O - C) behaviour of XTE~J1710-281, is also different from that seen in the aforementioned LMXBs.

The observed O - C variation strongly resemble the one seen in EXO 0748-676 \citep{Wolff09}. Interestingly, of the known LMXBS with well determined eclipse times, EXO 0748-676 and XTE J1710-281 have shortest duration of the eclipse, making it easier to monitor with X-ray observatories. Though fewer eclipses have been observed in XTE J1710-281, as opposed to more than 400 complete eclipses seen from EXO 0748-676 \citep{Wolff09}, the mid-eclipse times measured with RXTE-PCA are accurate enough to enable detection of very small orbital period glitches.

Magnetic field cycling of the secondary star is assumed to be the likely cause for the observed orbital period glitches in EXO 0748-676 \citep{Wolff09}. It is proposed that magnetic activity associated with the secondary star could be responsible for sudden changes in the orbital period. If the secondary star associated with XTE J1710-281 has strong, changing, magnetic activity, it can lead to changes in the structure of secondary star. A changing gravitational quadrapole moment can result into changes in the orbital period of the binary system \citep{Lanza99, Tauris06}. However, in case of XTE J1710-281, the optical counterpart has been discovered \citep{Ratti10} but the type of the companion star is not yet known. Therefore, it is difficult to make a statement on the probable cause for the changing orbital period of XTE J1710-281.

We emphasise that if magnetic cycling of the binary components is indeed a reason behind the observed epochs of orbital period, then long term monitoring of XTE J1710-281, is required to determine the timescales of magnetic cycling of the secondary star. It may also be useful to foretell the distinct orbital period epochs of XTE J1710-281, if any. Lastly, the forthcoming Indian-satellite, $ASTROSAT$ with a very large area X-ray proportional counter \citep{Paul09} could be a boon in determining the orbital parameters of the system. 
 
\section*{Acknowledgments}

This research has made use of data obtained from the High Energy Astrophysics Science Archive Research Center (HEASARC), provided by NASA's Goddard Space Flight Center.

\label{lastpage}

\end{document}